\documentclass[aps,prd,reprint,showpacs,longbibliography,groupedaddress,titlepage]{revtex4-1} 

\usepackage{amssymb}   
\usepackage{mathtools} 
\usepackage{hyperref}
\hypersetup{colorlinks=true,linkcolor=blue,urlcolor=blue,citecolor=blue}
\usepackage{accents}
\usepackage{tensor}
\usepackage[cal=boondox]{mathalfa}
\usepackage{enumerate}

\usepackage[T1]{fontenc}

\begin{document}

\title{Plebanski-like action for general relativity and anti-self-dual gravity}

\author{Mariano Celada}
\email[]{mcelada@fis.cinvestav.mx}
\affiliation{Departamento de F\'{\i}sica, Cinvestav, Instituto Polit\'ecnico Nacional 2508, San Pedro Zacatenco,
	07360, Gustavo A. Madero, Ciudad de M\'exico, M\'exico}

\author{Diego Gonz\'alez}
\email[]{dgonzalez@fis.cinvestav.mx}
\affiliation{Departamento de F\'{\i}sica, Cinvestav, Instituto Polit\'ecnico Nacional 2508, San Pedro Zacatenco,
	07360, Gustavo A. Madero, Ciudad de M\'exico, M\'exico}

\author{Merced Montesinos} 
\email[]{merced@fis.cinvestav.mx}
\affiliation{Departamento de F\'{\i}sica, Cinvestav, Instituto Polit\'ecnico Nacional 2508, San Pedro Zacatenco, 
	07360, Gustavo A. Madero, 
	Ciudad de M\'exico, M\'exico}

\date{\today}

\begin{abstract}
We present a new $BF$-type action for complex general relativity with or without a cosmological constant resembling Plebanski's action, which depends on an SO(3,$\mathbb{C}$) connection, a set of 2-forms, a symmetric matrix, and a 4-form. However, it differs from the Plebanski formulation in the way that the symmetric matrix enters into the action. The advantage of this fact is twofold. First, as compared to Plebanski's action, the symmetric matrix can now be integrated out, which leads to a pure $BF$-type action principle for general relativity; the canonical analysis of the new action then shows that it has the same phase space of the Ashtekar formalism up to a canonical transformation induced by a topological term. Second, a particular choice of the parameters involved in the formulation produces a $BF$-type action principle describing conformally anti-self-dual gravity. Therefore, the new action unifies both general relativity and anti-self-dual gravity.
\end{abstract}


\maketitle

\section{Introduction}\label{intro}

The Plebanski formulation~\cite{pleb1977118} has played a major role in understanding the classical and quantum aspects of general relativity (GR)~\cite{rovelli2004quantum,thiemann2007modern,perez2013-16-3}.  Having been the first $BF$-type formulation (even when the terminology $BF$ had not been invented yet) for Einstein's theory, it establishes that instead of the metric tensor, the fundamental variables to describe the gravitational field are (self-dual) 2-forms. More precisely, the variables involved in the Plebanski formulation are, besides the 2-forms ($B$ fields) and the gauge connection, a symmetric matrix $\Psi$ which plays the role of a Lagrange multiplier and that imposes the so-called simplicity constraint on the $B$ fields, and a 4-form imposing a constraint on $\text{Tr}\Psi$. It turns out that by following a systematic procedure in which some of these variables are integrated out, the pure connection formulation for GR is obtained \cite{CGM_2015}, which was originally introduced following a different approach \cite{kras2011106}. All of this shows that GR is a special type of diffeomorphism-invariant gauge theory. Since gauge theories admit a pure $BF$-type formulation with the gauge connection and the $B$ fields as the only fundamental variables \cite{El1973,Halpern.16.1798,Cattaneo}, it is desirable to have an analogous action principle for GR. In this line of thought, a new $BF$-type action principle for GR with a nonvanishing cosmological constant was reported in Ref.~\cite{Krasn2015}. Nevertheless, the origin of such an action principle is unclear since it cannot be obtained from the Plebanski formulation, which stems from the fact that it is not possible to integrate out $\Psi$ while keeping both the gauge connection and the $B$ fields. 

On the other side, conformally anti-self-dual gravity \cite{Capo-7-1-001,Koshti,Torreprd41,PhysRevD.50.7323} describes solutions to (Euclidean) Einstein's equations  (also known as gravitational instantons) with a vanishing self-dual Weyl curvature. As shown in Ref.~\cite{Koshti}, solutions of this type with a nonvanishing cosmological constant come from the Samuel ansatz~\cite{Samuel1988}, which is the covariant version of the Ashtekar-Renteln ansatz~\cite{AshlectNote}, introduced as a possible solution to the constraints of the Ashtekar formalism for GR. Later, this ansatz is traduced into a quadratic constraint on the self-dual curvature known as instanton equation~\cite{Capo-7-1-001}, for which an action functional involving 2-forms is given afterwards~\cite{torre1990topological}. However, the relation between this action and Plebanski's action has not been established yet. 

In this paper we present a new Plebanski-like action principle for GR  with or without a cosmological constant that (i) clarifies the origin of the $BF$-type action principle reported in Ref.~\cite{Krasn2015} and (ii) provides a $BF$-type action for anti-self-dual gravity which leads to the action reported in Ref.~\cite{torre1990topological}. It involves the same variables of Plebanski's action but has a different functional form. Remarkably, the new dependency makes the field $\Psi$ an auxiliary field from the very beginning and thus can be integrated out in the action before integrating out any other field. By doing so, we obtain a generalization of the $BF$-type action reported in Ref.~\cite{Krasn2015} that also holds for the case with a vanishing cosmological constant, and that reduces to the action of Ref.~\cite{Krasn2015} after integrating out the 4-form. Furthermore, for a particular choice of the parameters involved, the Plebanski-like action describes conformally anti-self-dual gravity, and, by eliminating the $B$ field and the 4-form from it, we make contact with the action reported in Ref.~\cite{torre1990topological}. In this sense, the action we present unifies both complex GR and conformally anti-self-dual gravity.

The outline of this paper is as follows. First, in Sec. \ref{sect2} we set up the new action principle and show that the equations of motion arising from it imply the Plebanski equations of motion for GR. Second, in Sec. \ref{sect3} we proceed to integrate out the auxiliary fields appearing in the action and show that the action of Ref.~\cite{Krasn2015} as well as the pure connection formulation of GR arise; particularly, we realize that the proposed action contains an intrinsic topological term. Afterwards, in Sec. \ref{sect4} we perform the Hamiltonian analysis of the proposed action and find that the phase space is described by the Ashtekar variables subject to the same constraints of GR up to a canonical transformation induced by the already mentioned topological term. Later, in Sec. \ref{sect2a} we show how the new action describes conformally anti-self-dual gravity. Finally, our conclusions are collected in Sec. \ref{concl}.

\section{ A Plebanski-like formulation}\label{sect2}

We start by setting up the aforementioned action principle. It its given by
\begin{eqnarray}
	& S[A,B,\rho,\Psi]=\int\bigl[B_i\wedge F^i\nonumber\\
	&+\frac{1}{2}\left(\Psi_{ij}-\lambda\delta_{ij}\right)B^i\wedge B^j+\left(\beta\text{Tr}\Psi^{-1}-\gamma\right)\rho\bigr].\label{actABrP}
\end{eqnarray}
Here $F^i:=d A^i + (1/2) \tensor{\varepsilon}{^i_{jk}}  A^j  \wedge A^k$ is the curvature of the SO$(3,\mathbb{C})$ connection $A^i$, $B^i$ are three $\mathfrak{so}(3,\mathbb{C})$-valued 2-forms, $\Psi_{ij}$ is a 3$\times$3 complex invertible symmetric matrix that imposes certain constraints on $B^i$, and $\rho$ is a nonvanishing complex-valued 4-form. The parameters $\lambda$, $\beta$, and $\gamma$ are complex in principle and are related to the cosmological constant. If the connection $A^i$ is dimensionless as a 1-form and $[\Lambda]$ stands for the dimension of the cosmological constant, then $\lambda$, $\beta$, and $\gamma$ have dimension $[\Lambda]$, $[\Lambda]^2$, and $[\Lambda]$, respectively (thus, the action has dimension $[\Lambda]^{-1}$). The group indices are raised and lowered with the Kronecker delta $\delta^{ij}$, and ${\varepsilon}_{ijk}$ is the Levi-Civita symbol  (${\varepsilon}_{123}=+1$).

The difference between the action (\ref{actABrP}) and Plebanski's action lies in the constraint imposed by the Lagrange multiplier $\rho$; while in Plebanski's action it imposes certain restriction on the trace of $\Psi$, here it imposes a similar restriction on the trace of $\Psi^{-1}$ instead, and this simple change allows us to integrate out $\Psi$ in the action (\ref{actABrP}) from the very beginning, which is not possible in the Plebanski formulation. Furthermore, after integrating out some of the auxiliary fields involved in~(\ref{actABrP}), we find that this change translates into the coupling of a topological term to gravity. Notice that the metric tensor is not involved in the action~(\ref{actABrP}), but once we identify the (self-dual) 2-forms that satisfy the simplicity constraint, it can be constructed from them by using Urbantke's formula~\cite{urba198425}.

Our claim is that the action principle (\ref{actABrP}) describes GR with or without a cosmological constant for nonvanishing $\lambda$ and $\beta$. To prove this, we will show that the equations of motion coming from (\ref{actABrP}) imply the Plebanski equations. The variation of (\ref{actABrP}) with respect to the independent variables leads to the following equations of motion:
\begin{subequations}
	\begin{eqnarray}
		&\delta A&:\ DB^i:=dB^i+\tensor{\varepsilon}{^i_{jk}}A^j\wedge B^k=0,\label{DB}\\
		&\delta \Psi&:\ B^i\wedge B^j-2\beta\rho(\Psi^{-1})^{ik}\tensor{(\Psi^{-1})}{^j_k}=0,\label{simplic}\\
		& \delta B&:\ F^i+(\tensor{\Psi}{^i_j}-\lambda\delta^i_j)B^j=0,\label{eqB}\\
		&\delta\rho&: \beta\text{Tr}\Psi^{-1}-\gamma=0\label{eqtr}.
	\end{eqnarray}
\end{subequations}
Let us assume that $\beta\neq0$ and define the three 2-forms $\Sigma^i$~by
\begin{equation}\label{sigmadef}
	\Sigma^i:=\beta^{-1/2}\tensor{\Psi}{^i_j}B^j.
\end{equation}
Then Eq. (\ref{simplic}) implies
\begin{equation}\label{simplsig}
	\Sigma^i\wedge\Sigma^j-2\rho\delta^{ij}=0,
\end{equation}
which means that $\Sigma^i$ satisfies the simplicity constraint involved in the Plebanski formulation. On the other hand, the Eqs. (\ref{DB}) and (\ref{eqB}), together with the Bianchi identity $DF^i=0$, imply that $\Sigma^i$ is covariantly constant, namely,
\begin{equation}\label{DSig}
	D\Sigma^i=0.
\end{equation}
Now we need to relate the 2-forms $\Sigma^i$ to the curvature $F^i$. Let us define the symmetric matrix $\Phi_{ij}$ by
\begin{equation}
	\Phi:=\lambda\beta^{-1/2}\left(\beta\Psi^{-1}-\frac{\gamma}{3}\text{Id}\right),\label{Xtraless}
\end{equation}
where Id is the $3\times3$ identity matrix. An immediate consequence of this definition and of Eq. (\ref{eqtr}) is 
\begin{equation}
	\text{Tr}\Phi=0\label{TrX=0}.
\end{equation}
By combining (\ref{eqB}), (\ref{sigmadef}), and (\ref{Xtraless}), we finally obtain
\begin{equation}
	F^i=\left(\tensor{\Phi}{^i_j}+\frac{1}{3}\Lambda\delta^i_j \right)\Sigma^i,\label{CurvSig}
\end{equation}
where $\Lambda:=\lambda\gamma\beta^{-1/2}-3\beta^{1/2}$ is the cosmological constant, which involves the three parameters introduced in the action principle (\ref{actABrP}). For $\lambda\neq 0$, Eqs. (\ref{simplsig}), (\ref{DSig}), (\ref{TrX=0}), and (\ref{CurvSig}) constitute the Plebanski equations of motion for GR with a cosmological constant, where $\Phi$ is identified as the self-dual part of the Weyl curvature.  Notice that $\gamma$ is required to allow a vanishing cosmological constant; in fact, for $\gamma=3\beta/\lambda$, $\Lambda$ vanishes. Therefore, we have shown that, for $\lambda\neq 0$ and $\beta\neq 0$, the action principle (\ref{actABrP}) describes GR with a vanishing or nonvanishing cosmological constant. The case $\lambda= 0$ and $\beta\neq 0$ also deserves to be mentioned, and it shall be analyzed below.

\section{Integrating out the auxiliary fields}\label{sect3}

We now move to one of the purposes of this paper, which is to show that the action principle (\ref{actABrP}) leads to the $BF$-type action reported in \cite{Krasn2015}. Let us start by integrating out $\Psi$ from the action (\ref{actABrP}). From Eq. (\ref{simplic}), the solution for $\Psi$ involves the square root of the matrix $\tilde{N}$ defined by $B^i\wedge B^j=:\tilde{N}^{ij}d^4x$. Square roots of matrices  always do exist for invertible matrices~\cite{horn1994topics}, although, in general, they are not unique. Since in our case we are dealing with invertible matrices, solutions to (\ref{simplic}) do indeed exist. Denoting by $\tilde{N}^{1/2}$ the (symmetric) square root of $\tilde{N}$, the solution of (\ref{simplic}) is
\begin{equation}
\Psi=\sqrt{2}\varepsilon\beta^{1/2}\tilde{\rho}^{1/2}\tilde{N}^{-1/2},\label{solvpsi}
\end{equation}
where $\tilde{N}^{-1/2}$ is the inverse of $\tilde{N}^{1/2}$, $\rho=:\tilde{\rho}d^4x$, and $\varepsilon=\pm1$ accounts for the branch of $\tilde{N}^{1/2}$. Notice that here we are taking one of the squares roots of $\tilde{N}$, but the procedure is valid regardless of the root chosen. Plugging this back into (\ref{actABrP}) yields
\begin{eqnarray}
& S[A,B,\rho]=\int\Bigl(B^i\wedge F^i\nonumber\\
&-\frac{\lambda}{2}B_i\wedge B^i+\sqrt{2}\varepsilon\beta^{1/2}\tilde{\rho}^{1/2}\text{Tr}\tilde{N}^{1/2}d^4 x-\gamma\rho\Bigr).\label{actABR}
\end{eqnarray}
This action principle, which cannot be obtained directly from the Plebanski formulation, offers us two possibilities: we can integrate out either the 2-forms $B^i$ or the field $\rho$. Since one of our aims is to arrive at the action principle of Ref. \cite{Krasn2015}, we proceed first to get rid of $\rho$, which requires $\gamma\neq0$. If we make that assumption, the equation of motion for $\rho$ implies
\begin{equation}
\tilde{\rho}=\frac{\beta}{2\gamma^2}(\text{Tr}\tilde{N}^{1/2})^2.\label{exprrho}
\end{equation}
Then, by substituting this back into (\ref{actABR}), we obtain
\begin{eqnarray}
& S[A,B]=\int\Bigl[B_i\wedge F^i\nonumber\\
& -\frac{\lambda}{2}B_i\wedge B^i+\frac{\beta}{2\gamma}(\text{Tr}\tilde{N}^{1/2})^2 d^4 x\Bigr].\label{actAB}
\end{eqnarray}

The action principle (\ref{actAB}) is, as promised, the one reported in Ref. \cite{Krasn2015}; hence, in this paper we have established in a clean fashion that (\ref{actAB}) emerges from the action (\ref{actABrP}) through the elimination of some auxiliary fields. We remind the reader that the action principle (\ref{actAB}) only addresses the case with a nonvanishing cosmological constant, which also requires $\gamma\neq3\beta/\lambda$ (see the details in \cite{Krasn2015}); in contrast, a remarkable property of the action principle (\ref{actABR}) is that it supports $\Lambda=0$ and $\Lambda \neq 0$ without involving the field $\Psi$.

Following the second option, we integrate out the 2-forms $B^i$ in (\ref{actABR}) instead, which allows us to keep $\gamma$ arbitrary. The equation of motion for the 2-forms $B^i$ yields
\begin{equation}
	F^i+\sqrt{2}\varepsilon\beta^{1/2}\tilde{\rho}^{1/2}\tensor{(\tilde{N}^{-1/2})}{^i_j}B^j-\lambda B^i=0,\label{EqB}
\end{equation}
whose solution for $B^i$ is
\begin{equation}
B^i=\frac{1}{\lambda}\left[\varepsilon\sqrt{2}\beta^{1/2}\tilde{\rho}^{1/2}\tensor{(\tilde{M}^{-1/2})}{^i_j}+\delta^i_j\right]F^j,\label{solB}
\end{equation}
where we have defined $F^i\wedge F^j=:\tilde{M}^{ij}d^4x$. Substituting this solution into the action (\ref{actABR}) yields 
\begin{eqnarray}
& S[A,\rho]=\frac{1}{2\lambda}\int F_i\wedge F^i\nonumber\\
&+\frac{\beta^{1/2}}{\lambda}\int d^4x\left[\sqrt{2}\varepsilon\tilde{\rho}^{1/2}\text{Tr}\tilde{M}^{1/2}-\Lambda\tilde{\rho}\right].\label{acAR}
\end{eqnarray}
The first term on the rhs of (\ref{acAR}) is topological and does not affect the classical dynamics, while the second term is the same term obtained after integrating out $\Sigma^i$ and $\Psi$ in the Plebanki's action and describes GR for both a vanishing and a nonvanishing cosmological constant \cite{CGM_2015} (see \cite{capo1991859} for a heuristic approach). Therefore, the action (\ref{acAR}) makes it clear that (\ref{actABrP}) contains an intrinsic topological term. In the case of $\Lambda\neq0$, the variable $\tilde{\rho}$ can be integrated out from (\ref{acAR}), and this leads to the pure connection description of general relativity (coupled to a topological term):
\begin{equation}
S[A]=\frac{1}{2\lambda}\int F_i\wedge F^i+\frac{\beta^{1/2}}{2\lambda\Lambda}\int d^4x \left(\text{Tr}\tilde{M}^{1/2}\right)^2.\label{pure1}
\end{equation}
It is worth pointing out that the action (\ref{pure1}) depends on the particular square root of $\tilde{M}$, but, because of the square in the second term on the rhs, it does not depend on the branch. Notice that, by eliminating the 2-forms $B^i$ from (\ref{actAB}), we also arrive at (\ref{pure1}).

One of the advantages of the action principle (\ref{actABrP}) is that it provides a way to not only get rid of $\Psi$ from the action, but also to eliminate $B^i$ from the very beginning. Let us now follow this second path. The equation of motion (\ref{eqB}) can be (uniquely) solved for $B^i$ only if $\det(\Psi-\lambda\text{Id})\neq 0$, which is, in general, satisfied. By substituting this solution into (\ref{actABrP}) we obtain
\begin{eqnarray}
	& S[A,\Psi,\rho]=\nonumber\\
	&\int\left[\frac{1}{2}(\chi^{-1})_{ij} F^i\wedge F^j+\left(\beta\text{Tr}\Psi^{-1}-\gamma\right)\rho\right],\label{acAPR}
\end{eqnarray}
where $\chi:=\lambda\text{Id}-\Psi$. Now we integrate out $\Psi$ in (\ref{acAPR}). The solution for $\Psi$ due to its own equation of motion takes the form 
\begin{equation}
	\Psi=\lambda\left(\text{Id}+\frac{\epsilon}{\sqrt{2}}\beta^{-1/2}\tilde{\rho}^{-1/2}\tilde{M}^{1/2}\right)^{-1},\label{Solv2}
\end{equation}
where $\epsilon=\pm1$ accounts for the ambiguity in the branch of $\tilde{M}^{1/2}$. Substituting this back into the action (\ref{acAPR}), we recover the action (\ref{acAR}), with $\epsilon$ taking the place of $\varepsilon$.

\section{Canonical analysis}\label{sect4}

We now perform the canonical analysis of the action~(\ref{actABrP}). To make things easier, it is convenient to integrate out the variable $\Psi$ from the action, which leads to the equivalent action (\ref{actABR}). We assume that the spacetime has the topology $\mathbb{R}\times\Omega$, where $\Omega$ is a compact spatial 3-manifold without a boundary. We denote the spatial indices by $a,\ b,\dots=1,2,3$, while the time component is referred to as the 0-component. The 3+1 decomposition of the action (\ref{actABR}) yields
\begin{eqnarray}
	& S[A,B,\rho]=\int_{\mathbb{R}}dt\int_{\Omega}d^3x\Bigr[\tilde{\Pi}^{ai}\dot{A}_{ai}+A_{0i}\tilde{\mathcal{G}}^i\nonumber\\
	&+B_{0ai}\tilde{E}^{ai}+\sqrt{2}\varepsilon\beta^{1/2}\tilde{\rho}^{1/2}\text{Tr}\tilde{N}^{1/2}-\gamma\tilde{\rho}\Bigr],\label{Act3+1}
\end{eqnarray}
where we have defined $\tilde{\Pi}^{ai}:=(1/2)\tilde{\eta}^{abc}\tensor{B}{_{bc}^i}$, $\tilde{E}^{ai}:=\tilde{B}^{ai}-\lambda\tilde{\Pi}^{ai}$ for $\tilde{B}^{ai}:=(1/2)\tilde{\eta}^{abc}\tensor{F}{_{bc}^i}$, $\tilde{\mathcal{G}}^i:=\mathcal{D}_a\tilde{\Pi}^{ai}$, with $\mathcal{D}_a$ being the SO(3,$\mathbb{C}$)-covariant derivative and $\tilde{\eta}^{abc}$ $(\underaccent{\tilde}{\eta}_{abc})$ the Levi-Civita symbol of weight 1 (-1). 

Since the action (\ref{Act3+1}) depends on $\tilde{N}^{1/2}$, it is nonpolynomial in the variable $B_{0ai}$. However, this variable is not dynamical because no time derivatives of it appear in (\ref{Act3+1}). Hence, we can use the equation of motion corresponding to $B_{0ai}$ to eliminate it from the action. The variation of (\ref{Act3+1}) with respect to $B_{0ai}$ leads to
\begin{equation}
	\tilde{E}^{ai}+\sqrt{2}\varepsilon\beta^{1/2}\tensor{(\varphi^{-1})}{^i_j}\tilde{\Pi}^{aj}=0,\label{consEP}
\end{equation}
where the matrix $\varphi_{ij}:=\tilde{\rho}^{-1/2}(\tilde{N}^{1/2})_{ij}$ is symmetric. This equation must be solved for $B_{0ai}$. Assuming that $\det(\tilde{\Pi}^{ai})\neq 0$ , then Eq. (\ref{consEP}) allows us to express $\varphi$ in terms of the variables $A_{ai}$ and $\tilde{\Pi}^{ai}$, which constitute the phase space's canonical pair according to the first term on the rhs of (\ref{Act3+1}). On the other hand, we can show from the definitions of the matrices $\tilde{N}$ and $\varphi$ that the expression $\tensor{X}{_a^i}\tilde{Y}^{aj}+\tensor{X}{_a^j}\tilde{Y}^{ai}-\tilde{\rho}\delta^{ij}=0$, with $X_{ai}:=(\varphi^{-1})_{ij}\tensor{B}{_{0a}^j}$ and $\tilde{Y}^{ai}:=\tensor{(\varphi^{-1})}{^i_j}\tilde{\Pi}^{aj}$, is satisfied. This last equation can be solved for $X_{ai}$ as $X_{ai}=\left[(\tilde{\rho}/2)\delta_{ij}+\varepsilon_{ijk}\tilde{N}^k\right]\tensor{\underaccent{\tilde}{Y}}{_a^j}$, where $\underaccent{\tilde}{Y}_{ai}$ is the inverse of $\tilde{Y}^{ai}$, and $\tilde{N}^k$ is an arbitrary internal three-vector of weight 1. Using this and (\ref{consEP}), and denoting by $\underaccent{\tilde}{E}_{ai}$ the inverse of $\tilde{E}^{ai}$, the solution for $B_{0ai}$ takes the form
\begin{equation}
	\hspace{-2mm}B_{0ai}=2\beta\left[\frac{1}{2}\tilde{\rho}\underaccent{\tilde}{E}_{cj}\tensor{\underaccent{\tilde}{E}}{_a^j}+(\det\tilde{E})^{-1}\underaccent{\tilde}{\eta}_{abc}\tensor{\tilde{E}}{^b_j}\tilde{N}^j\right]\tensor{\tilde{\Pi}}{^c_i},\label{solB_0a}
\end{equation}
with $\det\tilde{E}:=\det(\tilde{E}^{ai})$. Plugging (\ref{solB_0a}) back into (\ref{Act3+1}), the action becomes
\begin{eqnarray}
	& S[A_{ai},\tilde{\Pi}^{ai},A_{0i},N^a,\underaccent{\tilde}{N}]=\nonumber\\
	&\int_{\mathbb{R}}dt\int_{\Omega}d^3x\left(\tilde{\Pi}_{ai}\dot{A}^{ai}+A_{0i}\tilde{\mathcal{G}}^i+N^a\tilde{\mathcal{V}}_a+\underaccent{\tilde}{N}\tilde{\tilde{\mathcal{H}}}\right),\label{actham}
\end{eqnarray}
where $N^a:=2\beta(\det\tilde{E})^{-1}\tensor{\tilde{E}}{^a_i}\tilde{N}^i$ and $\underaccent{\tilde}{N}:=-(\det\tilde{E})^{-1}\tilde{\rho}$. Since $A_{0i}$, $N^a$, and $\underaccent{\tilde}{N}$ are arbitrary and appear linearly in the action, they play the role of Lagrange multipliers and impose the following constraints:
\begin{subequations}
	\begin{eqnarray}
		&\mathcal{\tilde{G}}^i=\mathcal{D}_a\tilde{\Pi}^{ai}\approx 0,\label{gauss}\\
		&\mathcal{\tilde{V}}_a:=\tilde{\Pi}^{bi}F_{bai}\approx 0,\label{Vect}\\
		&\tilde{\tilde{\mathcal{H}}}:=\gamma BBB-3(\gamma\lambda-\beta)\Pi B B\nonumber\\
		&+3\lambda(\gamma\lambda-2\beta)\Pi\Pi B-\lambda^2(\gamma\lambda-3\beta)\Pi\Pi\Pi\approx 0,\label{scalar}
	\end{eqnarray}
\end{subequations}
where we have used the shorthand $\Pi\Pi B:=(1/6)\underaccent{\tilde}{\eta}_{abc}\varepsilon_{ijk}\tilde{\Pi}^{ai}\tilde{\Pi}^{bj}\tilde{B}^{ck}$, etc. It turns out that the constraints (\ref{gauss})--(\ref{scalar}) are the only ones of the theory and that they are first class despite the complicated form of $\tilde{\tilde{\mathcal{H}}}$ \cite{beng1991254}, which implies that they generate the gauge symmetries of the theory: the Gauss constraint $\mathcal{\tilde{G}}^i$ generates local SO(3,$\mathbb{C}$) rotations, while the vector and scalar constraints ($\mathcal{\tilde{V}}_a$ and $\tilde{\tilde{\mathcal{H}}}$, respectively) generate spacetime diffeomorphisms. Since we have nine configuration variables $A_{ai}$ and seven first-class constraints, the number of physical (complex) degrees of freedom per space point is two, as expected from a theory describing GR. Notice that the constraint (\ref{scalar}) does not have the same form as the scalar constraint of the Ashtekar formalism, where only the terms $\Pi\Pi B$ and $\Pi\Pi\Pi$ show up (the latter only for a nonvanishing cosmological constant); in fact, the terms $BBB$ and $\Pi BB$ are the result of the presence of an intrinsic topological term in action (\ref{actABrP}) [which explicitly shows up in (\ref{acAR}) after integrating out the auxiliary fields $B$ and $\Psi$ in (\ref{actABrP})], but we have the freedom to perform the canonical transformation $(A_{ai},\tilde{\Pi}^{ai})\rightarrow(A_{ai},\tilde{\Pi}^{ai}+\theta \tilde{B}^{ai})$ to cancel the effect of this topological term and bring the scalar constraint (\ref{scalar}) to the form of the Ashtekar one \cite{ashtekar1989-04-06-1493,beng1991254}.

Alternatively, one also can read the vector and scalar constraints from the equation of motion (\ref{consEP}). Indeed, the fact that the matrix $\varphi$ is symmetric gives rise to the vector constraint $\tilde{\mathcal{V}}_a$, while the scalar constraint $\tilde{\tilde{\mathcal{H}}}$ results from the combination of $\text{Tr}\varphi$ from (\ref{consEP}) with the equation of motion corresponding to $\tilde{\rho}$ from (\ref{Act3+1}).


\section{Anti-self-dual gravity}\label{sect2a}

Let us now consider the case with $\lambda=0$ and $\beta\neq0$ in the action~(\ref{actABrP}). Here, the equations of motion for $A$,  $\Psi$, and $\rho$ remain unchanged, whereas the equation for $B$ yields
\begin{eqnarray}
F^i+\tensor{\Psi}{^i_j}B^j=0.\label{eqB2}
\end{eqnarray}
By introducing the definition~(\ref{sigmadef}), we still have (\ref{simplsig}) as a consequence of~(\ref{simplic}). Using~(\ref{sigmadef}) and (\ref{eqB2}), we obtain
\begin{eqnarray}
F^i=\frac{\Lambda}{3}\Sigma^i, \label{eqF}
\end{eqnarray}
where $\Lambda=-3\beta^{-1/2}$ is the cosmological constant. Notice that, in Plebanski's formulation, the curvature and the 2-forms satisfying the simplicity constraint are linearly related via a matrix of the form $W^i{}_j+(\Lambda/3)\delta^i_j$, where $W^i{}_j$ is the self-dual Weyl curvature. In view of this, Eq. (\ref{eqF}) has as an immediate consequence that $W^i{}_j$ vanishes; hence, the Weyl tensor is purely anti-self-dual. 

By combining (\ref{simplsig}) and (\ref{eqF}), we obtain the instanton equation~\cite{Torreprd41,Capo-7-1-001}
\begin{eqnarray}
F^i\wedge F^j-\frac{1}{3}\delta^{ij}F^k\wedge F_k=0,\label{instanton}
\end{eqnarray}
which characterizes anti-self-dual gravitational instantons. Notice that due to both Bianchi's identity $DF^i=0$ and Eq. (\ref{instanton}), the internal connection is still the self-dual part of the spin connection.   
Therefore, we conclude that for $\lambda=0$ and $\beta\neq0$ the action (\ref{actABrP}) describes conformally anti-self-dual gravity.

We point out that after eliminating $\rho$ from the action (\ref{acAPR}) with $\lambda=0$ and making the change $X:=-\Psi^{-1}$, the resulting action is
\begin{eqnarray}
\hspace{-2mm} S[A,X]=\frac{1}{2}\int\left[(X_{\rm tf})_{ij} F^i\wedge F^j - \frac{\gamma}{3 \beta} F^i\wedge F_i\right],
\end{eqnarray}
where $X_{\rm tf}$ is the trace-free part of $X$. This action corresponds (up to the topological term) to the action introduced in Ref. \cite{torre1990topological} to describe the moduli space of anti-self-dual gravitational instantons. This fact strengthens the ability of action (\ref{actABrP}) for supporting anti-self-dual gravity.

Moreover, in this case the Gauss and vector constraints remain unchanged, while the scalar constraint takes the form
\begin{equation}
	\tilde{\tilde{\mathcal{H}}}:=\gamma BBB+3\beta\Pi B B\approx 0,\label{scal_self}	
\end{equation}
which is dual to the one of the Ashtekar formalism in the sense that it results from the Ashtekar scalar constraint (with a cosmological constant) after interchanging $\tilde{\Pi}^{ai}$ and $\tilde{B}^{ai}$. However, the first term on the rhs of (\ref{scal_self}) can be canceled by performing the canonical transformation $\tilde{\Pi}^{ai}\rightarrow\tilde{\Pi}^{ai}-(\gamma/3\beta)\tilde{B}^{ai}$, which leaves the Gauss and vector constraints invariant; the scalar constraint for anti-self-dual gravity then reads
\begin{equation}
\tilde{\tilde{\mathcal{H}}}=\Pi B B\approx 0,\label{scal_self_!}	
\end{equation}
where the constant factor has been dropped. This constraint cannot be mapped to the Ashtekar scalar constraint by using the aforementioned canonical transformation.



\section{Conclusions}\label{concl}

We conclude this paper by making some remarks.

\begin{enumerate}[(i)]
	\item We have presented a Plebanski-like action principle (\ref{actABrP})  for GR  with a vanishing or nonvanishing cosmological constant. It allows us to obtain the $BF$-type action (\ref{actAB}) of Ref. \cite{Krasn2015} in a clean fashion, which is not possible starting from Plebanski's action. We also show that by eliminating the auxiliary fields in the action (\ref{actABrP}), it is possible to obtain the pure connection action (\ref{pure1}), pointing out that the action (\ref{actABrP}) contains an intrinsic topological term. Finally, the canonical analysis of the action (\ref{actABrP}) leads to the phase space of the Ashtekar formulation of GR up to a canonical transformation.
	\item A consequence of having a functional dependence on the variables different from that of the Plebanski formulation is that the geometrical meaning of the variables changes. Now the $B$ fields do not satisfy the simplicity constraints of the Plebanski formulation, which means that we need to identify the 2-forms that do  [see Eq.~(\ref{sigmadef})]; the Urbantke metric is then constructed from them. Also, the matrix $\Psi$ is no longer the self-dual part of the Weyl tensor, but $\Psi^{-1}$ becomes a shifted self-dual part of it [see Eq. (\ref{Xtraless})].
	\item In the context of the formulations explored in Ref.~\cite{gm201591024021}, the action principle (\ref{acAR}) can alternatively be written as $S[A,\rho]=\int \rho[(1/\lambda) \text{Tr}\psi^2-\gamma ]$, where $\psi:=\tilde{\rho}^{-1/2}\tilde{M}^{1/2}/\sqrt{2}+\varepsilon \beta^{1/2} \text{Id}$. This indicates that GR can be obtained from the invariant $\text{Tr}\psi^2$ instead of $\text{Tr}\psi$.
	\item For $\lambda=0$ and $\beta\neq0$, the action (\ref{actABrP}) describes conformally anti-self-dual gravity. In fact, by eliminating the $B$ field and $\rho$ from it, we arrive at the action of Ref.~\cite{torre1990topological}, which encodes the dynamics of this theory.
	\item The case $\beta=0$ and $\lambda$ arbitrary is worth mentioning since, for that choice, the action (\ref{actABrP}) describes the Husain-Kuchar model~\cite{HK_mod}, which has a canonical structure similar to that of GR but lacks the Hamiltonian constraint.
	\item The Plebanski-like action (\ref{actABrP}), being of the $BF$ type, could be used as the starting point of the so-called spin foam models \cite{perez2013-16-3}, providing new insights into the covariant quantization of the gravitational field.
\end{enumerate}

\section*{ACKNOWLEDGMENTS}

This work was supported in part by Consejo Nacional de Ciencia y Tecnolog\'ia  (CONACyT), M\'exico, Grant No. 167477-F.

\bibliography{references}

\end{document}